\documentclass[10pt,a4paper,onecolumn]{article}
\usepackage[T1]{fontenc}
\usepackage{inputenc}
\usepackage[english]{babel}
\usepackage[nohead,includefoot,margin=2cm]{geometry}
\usepackage[compact,raggedright,small]{titlesec}
\usepackage[affil-it]{authblk}
\usepackage[runin]{abstract}
\usepackage{graphicx}
\usepackage{txfonts}
\usepackage{color}

\addto{\Affilfont}{\small}

\title{\large\bfseries Influence of Ta insertions on the magnetic properties of MgO/CoFeB/MgO films probed by ferromagnetic resonance}
\date{\small (Dated: \today)}
\author{Maria Patricia Rouelli Sabino}
\author{Sze Ter Lim}
\author{Michael Tran}
\affil{Data Storage Institute, Agency for Science, Technology and Research, 5 Engineering Drive 1, 117608 Singapore}

\begin{document}
\maketitle

\begin{abstract}
We show by vector network analyzer ferromagnetic resonance measurements that low Gilbert damping $\alpha$, down to 0.006, can be achieved in perpendicularly magnetized MgO/CoFeB/MgO thin films with ultrathin insertions of Ta in the CoFeB layer. Although increasing the number of Ta insertions allows thicker CoFeB layers to remain perpendicular, the effective areal magnetic anisotropy does not improve with more insertions, which come with an increase in $\alpha$. 
\end{abstract}

Perpendicular magnetic anisotropy (PMA) is the key to further downscaling of spin transfer torque magnetoresistive random memory devices, as it allows two key requirements to be satisfied: low critical current $I_{c0}$ and high thermal stability $\Delta$, the latter of which is proportional to the energy barrier $E_b$ between the two stable magnetic states. The spin torque switching efficiency, defined as $E_b/I_{c0}$, is commonly used as a metric to account for both requirements. For a Stoner-Wohlfarth model, it is given by \cite{Sun2013} $(\hbar/4e)\cdot(\eta/\alpha)$, where $\alpha$ is the Gilbert damping parameter, and $\eta$ is the spin polarization factor, which is related to the tunnel magnetoresistance ratio (TMR) by $\eta = [\mathrm{TMR}(\mathrm{TMR}+2)]^{1/2}/[2(\mathrm{TMR}+1)]$. It thus becomes evident that for high switching efficiency, one has to decrease $\alpha$ while keeping TMR high. Magnetic tunnel junctions (MTJs) based on CoFeB/MgO systems are well known to provide high TMR \cite{Yuasa2008} and have recently been shown to possess PMA, which is attributed to the CoFeB/MgO interface.\cite{Ikeda2010} A Ta layer is usually placed adjacent to the CoFeB to induce the proper crystallization necessary for PMA and high TMR \cite{D.C.Worledge2011}. In Ta/CoFeB/MgO systems, however, spin pumping to the Ta increases $\alpha$. \cite{Liu2011b} Moreover, the CoFeB layer also needs to be ultrathin (typically less than 1.5nm) in order to exhibit PMA. \cite{Ikeda2010} To improve the thermal stability as devices are scaled down to smaller diameters, increasing the effective areal anisotropy energy density $K_{eff}t$ is desired. 

One approach to address these issues is the use of double-MgO structures, i.e., those in which both the barrier layer and capping layer straddling the free layer are made of MgO. Improved $I_{c0}$ and/or $\Delta$ have been reported in devices using double MgO free layers. \cite{A2012,Park2012,Kubota2012,Sato2012} The improvement in thermal stability is attributed to the additional CoFeB/MgO interface, whereas lower $I_{c0}$ is associated with low $\alpha$. Indeed, $\alpha$ down to 0.005 has been measured in in-plane MgO/FeB/MgO films,\cite{Konoto2013} which agrees with device measurements.\cite{Tsunegi1882} The stacks investigated in these damping studies, however, did not have the Ta layer used in practical free layers with perpendicular anisotropy.\cite{Sato2012,Yakushiji2013} In addition, although the interfacial anisotropy in the out-of-plane devices measured by Tsunegi et al.\cite{Tsunegi1882} can be as high as 3.3 mJ/m$^2$, the effective perpendicular anisotropy was rather low ($K_{eff}t \approx 0.04$mJ/m$^2$) relative to that of a Ta/CoFeB/MgO stack\cite{Ikeda2010}. In this work, we explore the influence of Ta insertions within the CoFeB layer of MgO/CoFeB/MgO films by magnetometry and vector network analyzer ferromagnetic resonance (VNA-FMR) measurements. The insertion of extremely thin Ta layers (0.3 nm) inside the CoFeB layer aids crystallization, allowing a larger total CoFeB thickness to remain perpendicular, \cite{Naik2012} with an effective areal anistropy comparable to that of Ta/CoFeB/MgO.

Two sample series were deposited by magnetron sputtering on SiO$\mathrm{_2}$ substrates with seed layers of Ta 5/TaN 20/Ta 5 in an ultrahigh vacuum environment (all thicknesses in nm). The stack configurations of the two sample series are: (1) MgO 3/CoFeB 1.0/Ta 0.3/CoFeB 0.5 - 1.5/MgO 3 (``single-insertion'') and (2) MgO 3/CoFeB 1.0/Ta 0.3/CoFeB 0.5 - 1.5/Ta 0.3/CoFeB 1.0/MgO 3 (``double-insertion''), where the CoFeB composition is $\mathrm{Co_{40}Fe_{40}B_{20}}$ (at\%). The Ta insertion layer thickness is in the regime allowing strong ferromagnetic coupling between the CoFeB layers. \cite{Sokalski2012} Two other sample series were grown as references: (a) MgO 3/CoFeB 1.0 - 2.5/MgO 3 (``zero-insertion''), and (b) seed/CoFeB 1.0 - 2.5/MgO 3 (``single-MgO''). For all the double-MgO samples, an ultrathin CoFeB layer below the bottom MgO layer was also deposited for good MgO growth. We confirmed from separate measurements that this layer does not contribute to the magnetic signal.   All samples were capped with 15 nm of Ta for protection and were annealed post-growth at 300$^{\circ}$C for 1 h in vacuum.  Although 3 nm MgO is too thick for practical use in MTJs, it was chosen to ensure continuity of the MgO layers and lessen the influence of the layers beyond it.\cite{Konoto2013,Lam2013,Sabino2014} (Measurements of similar samples with 1 nm of MgO on both sides of the magnetic layer yielded the same trends.) 

\begin{figure}[hbt]
\begin{center}
\includegraphics[width=0.5\linewidth]{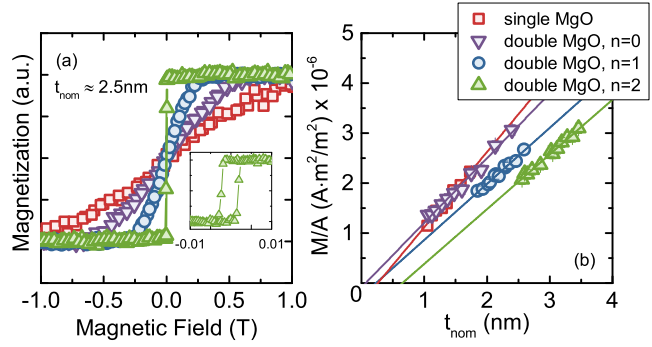}
\end{center}
\caption{(Color online) (a) Out-of-plane AGM loops for samples with single-MgO (red squares), zero-insertion (purple inverted triangles), single-insertion (blue circles), and double-insertion (green triangles), with total nominal CoFeB thickness $t_{nom} \approx$ 2.5 nm. Inset shows a low-field out-of-plane loop for the same double insertion sample. (b) Magnetic moment per unit area ($M/A$) as a function of the total nominal CoFeB thickness for all sample series. Linear fits are shown as solid lines. $M_S$ and $t_{MDL}$ can be extracted from the slope and $x$ intercept, respectively, and are summarized in Table \ref{tbl:summary}.}
\label{fig:agm}
\end{figure}

Magnetization measurements were performed using an alternating gradient magnetometer (AGM). The PMA improves with doubling of the CoFeB/MgO interface and with increasing number of Ta insertions, $n$, as shown in Fig.~\ref{fig:agm}(a) for samples with a similar total nominal CoFeB thickness $t_{nom} \approx$  2.5 nm. We also confirmed that we cannot obtain a perpendicular easy axis in double-MgO structures without Ta insertions.\cite{Sato2013} The double-insertion sample, on the other hand, exhibits large out-of-plane remanence as shown in the inset of Fig. \ref{fig:agm}(a). A coercive field less than 0.01 T (inset) is typical of CoFeB films with PMA\cite{Malinowski2009,Sokalski2012}. 

It is known that Ta can create a magnetically dead layer (MDL) when it is near a magnetic layer.\cite{Jang2011} We plot the magnetic moment per area  against $t_{nom}$ [Fig.~\ref{fig:agm}(b)] to obtain the thickness of the MDL for each series from the $x$ intercept of a linear fit. The results are summarized in Table \ref{tbl:summary}, along with the $M_S$ values obtained from the slope. We find an MDL thickness $t_{MDL}$ of $0.24\pm 0.09$ nm for $n = 1$ and $0.7 \pm 0.1$ nm for $n = 2$. These thicknesses are similar to the total Ta insertion thickness in the respective series, and are consistent with the picture of CoFeB intermixing with Ta to produce a magnetically dead volume. The $t_{MDL}$ value for the single-MgO samples (0.26 $\pm$ 0.08 nm) agrees with values found in the literature.\cite{Lam2013,Sinha2013a,Wang2006b} On the other hand, no dead layer was found for the zero-insertion samples, which is similar to the results in Ref. \cite{Jang2011}. 

	\begin{table*}[ht]
	\centering
	\small
		\caption{\ Summary of Magnetic Properties}
		\label{tbl:summary}
		\begin{tabular*}{0.8\textwidth}{@{\extracolsep{\fill}}lcccc}
			\hline
			Series					&	$t_{MDL}$ (nm)	&	$M_S$ (MA/m)	& $K_i \mathrm{(mJ/m^2)}$ 	& $K_v \mathrm{(MJ/m^3)}$ 		 \\
			\hline
			single MgO 			 &	$0.26 \pm 0.08$	&	$1.51 \pm 0.08$ & $1.61 \pm 0.07$	&	 $-0.29 \pm 0.08$	\\
			double MgO, $n$ = 0 & $0.04 \pm 0.06$	&	$1.29 \pm 0.05$ &	$0.91 \pm 0.09$	&	 $0.37  \pm 0.05$	\\
			double MgO, $n$ = 1 & $0.24 \pm 0.09$	& $1.12 \pm 0.05$ &	$2.18 \pm 0.08$	&	 $-0.34 \pm 0.04$	\\
			double MgO, $n$ = 2 & $0.7  \pm 0.1$	&	$1.10 \pm 0.04$	&	$2.4  \pm 0.1$	&	 $-0.25 \pm 0.03$ \\
			\hline
		\end{tabular*}
	\end{table*}

\begin{figure}[hb]
\begin{center}
\includegraphics[width=0.5\linewidth]{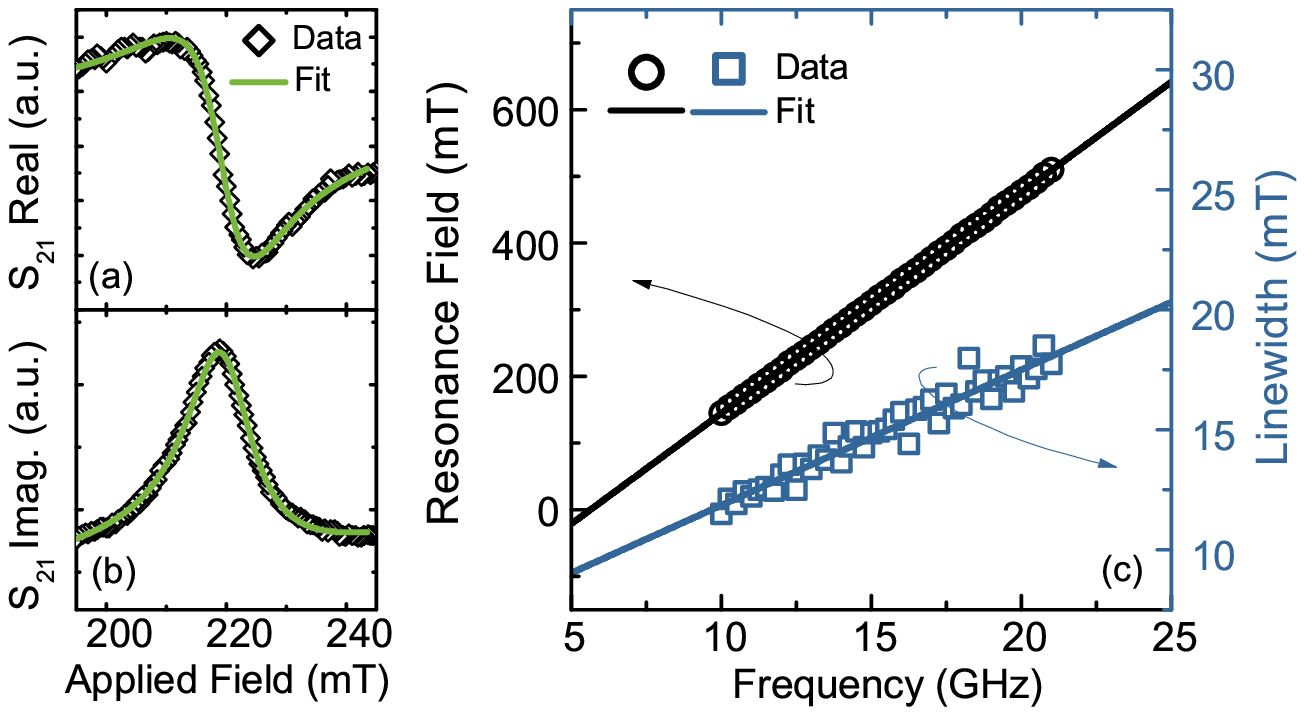}
\end{center}
\caption{(a) Real and (b) imaginary parts of the $S_{21}$ parameter obtained from VNA-FMR measurements for a double-insertion sample with $t_{nom} = 2.5$ nm at 12 GHz while a perpendicular dc magnetic field is swept. The lines are fits to an expression using Eq.~\ref{eq:chi}, taking nonmagnetic contributions to $S_{21}$ and a linear drift into account. (c) Field-swept linewidth and resonance fields for the same sample as a function of frequency. The linear fits described in the text are used to extract $H_{Keff}(= -M_{eff})$ and $\alpha$.}
\label{fig:samplespectrum}
\end{figure}

VNA-FMR was used to measure the effective anisotropy field and damping parameter of the samples. In the VNA-FMR setup, the samples were placed face down on a coplanar waveguide and situated in a dc magnetic field of up to $1.2$ T applied perpendicular to the film plane. The transmission scattering parameter $S_{21}$ was measured at a specific frequency while the dc field was swept. For each sweep, the real and imaginary parts of the resonance response were fitted simultaneously using the complex susceptibility equation
\begin{equation}
\chi(H) = \frac{M_{eff} (H - M_{eff} + i \frac{\Delta H}{2})}{(H - M_{eff})^2 - \left(\frac{2 \pi \hbar f}{g \mu_B}\right)^2 + i\Delta H (H - M_{eff})}
\label{eq:chi}
\end{equation}
where $f$ is the frequency of the ac field, $M_{eff} = M_S - H_K^{\;\:\bot}$, $\Delta H$ is the full width at half-maximum, $H_K^{\;\:\bot}$ is the anisotropy field perpendicular to the plane, $g$ is the spectroscopic splitting factor, $\mu_B$ is the Bohr magneton, and $\hbar$ is the reduced Planck's constant. Nonmagnetic contributions to the $S_{21}$ parameter and a linear time-dependent drift of the instruments were taken into account during the fit. We note that only one resonance peak is observed within the range studied. A representative fit of the susceptibility data is shown in Fig.~\ref{fig:samplespectrum}(a) for a double-insertion sample with $t_{nom} = 2.5$ nm. In using Eq.~\ref{eq:chi}, a value of $g = 2$ is first assumed to obtain values for $M_{eff}$ and $\Delta H$, which does not affect the final result.

For each frequency, a resonance field
\begin{equation}
\mu_0H_{res}(f) = \frac{2\pi \hbar}{g \mu_B}f + \mu_0M_{eff}
\label{eq:Kittel}
\end{equation} 
according to Kittel's equation is calculated and plotted against the frequency, as shown in Fig.~\ref{fig:samplespectrum}(c). A linear fit, now with $g$ and $M_{eff}$ as fitting parameters, is then performed.  The effective anisotropy energy density $K_{eff}$ can be calculated from the effective anisotropy field $H_{Keff} (= -M_{eff})$ as $K_{eff} = H_{Keff}M_{S}/2$, noting that a positive anisotropy constant corresponds to a perpendicular easy axis. 

To obtain $\alpha$, we perform a linear fit of the measured FMR linewidth as a function of the frequency to
\begin{equation}
\mu_0\Delta H(f) = \frac{4\pi \hbar \alpha}{g \mu_B} f + \mu_0\Delta H_0 
\label{eq:linewidth}
\end{equation}
where $\Delta H_0$ is the inhomogeneous linewidth broadening, and the value of $g$ used is the fitted value from Eq.~\ref{eq:Kittel}.  We note that two-magnon scattering contributions to the linewidth are eliminated owing to the perpendicular measurement configuration\cite{Mo2008}. Such a fit is shown in Fig.~\ref{fig:samplespectrum}(c). Only data points taken well beyond the saturation field for each sample were used in the fit, and asymptotic analysis as described in Ref. \cite{Shaw2013a} for the accessible frequency range was also performed. 

\begin{figure}[hbt]
\begin{center}
\includegraphics[width=0.35\linewidth]{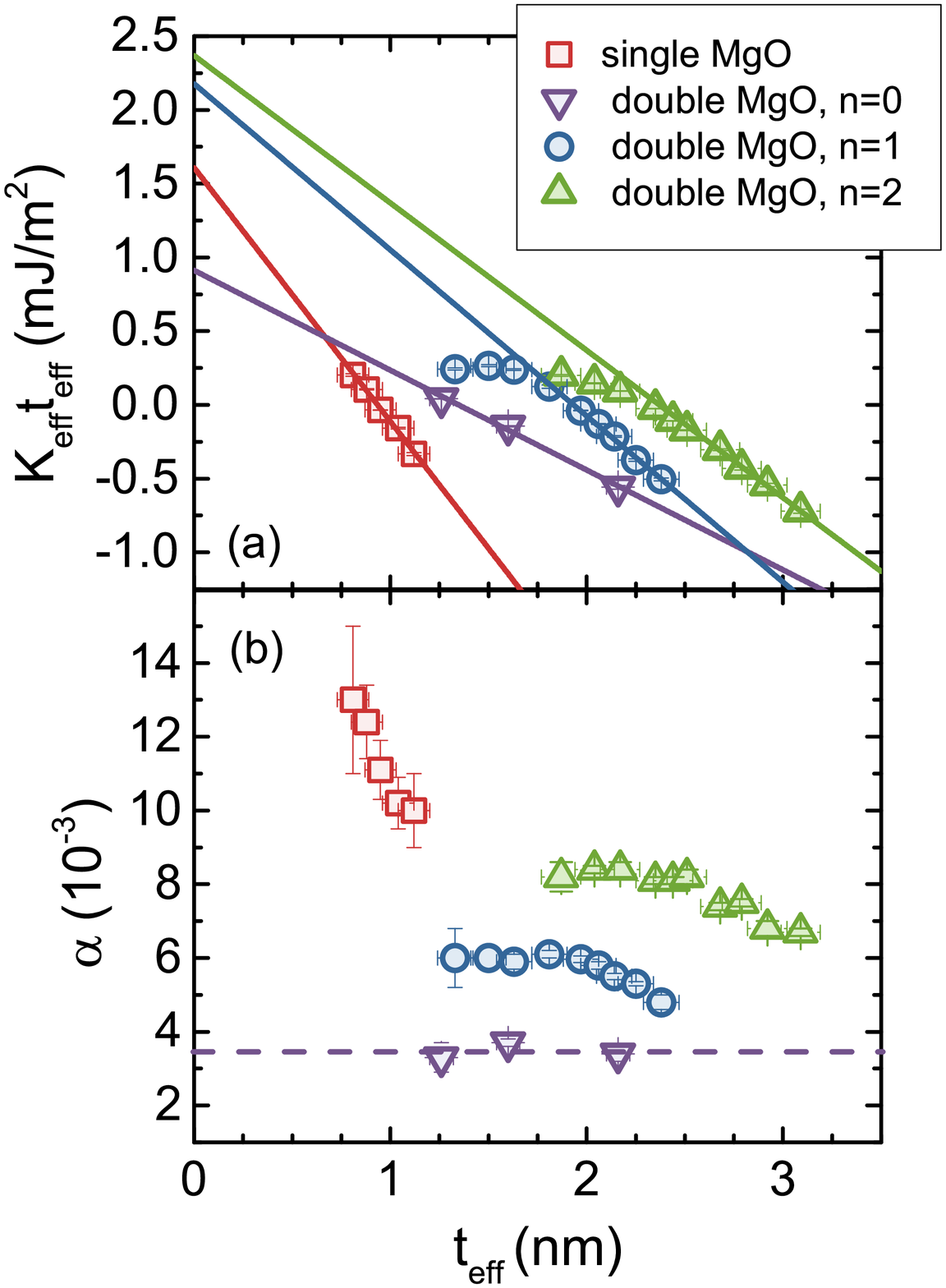}
\end{center}
\caption{(a)$K_{eff}t_{eff}$ and (b) $\alpha$ versus effective CoFeB thickness $t_{eff}$ obtained from field-swept VNA-FMR measurements for all sample series. Solid lines in (a) are linear fits. Purple dashed line in (b) corresponds to the mean $\alpha$ value averaged over all zero-insertion samples, which was found to be constant within error across the entire thickness range studied.} 
\label{fig:Keffalpha} 
\end{figure}

We define an effective thickness $t_{eff} = t_{nom} - t_{MDL}$ and show the calculated $K_{eff}t_{eff}$ (to which $E_b$ is proportional) for both sample series in Fig.~\ref{fig:Keffalpha}(a). The $x$-axis error bars originate from the fitting error in obtaining $t_{MDL}$. We find that for $t_{eff} > 2$ nm, double-insertion samples have higher $K_{eff}t_{eff}$ than single-insertion samples for the same $t_{eff}$. However, the maximum $K_{eff}t_{eff}$ achieved for both the single- and double-insertion series does not significantly exceed $K_{eff}t_{eff}$ measured in our thinnest single-MgO sample ($t_{eff}$ = 1.0 nm), similar to that observed in MTJ measurements.\cite{Park2012}

To understand this further, we consider the different contributions to $K_{eff}t_{eff}$, which is given by
\begin{equation}
K_{eff}t_{eff} = K_i + (K_v - \frac{\mu_0M_{S}^{2}}{2})t_{eff}
\label{eq:Keff}
\end{equation}
where  $K_i$ is the total interfacial anisotropy constant, including all CoFeB/MgO interfaces; $K_v$ is the volume anisotropy constant; and the demagnetizing energy is given by the $M_S^2$ term. We assume that any interfacial anisotropy from the Ta/CoFeB interface is negligible.\cite{Miura2013} $K_i$ is commonly derived from the $y$ intercept of a linear $K_{eff}t_{eff}$ versus $t_{eff}$ fit, whereas $K_v$ can be calculated from the slope if $M_S$ is known. Because it is possible that for CoFeB thicknesses below 1.0 nm, $K_i$ is degraded because of Ta reaching the CoFeB/MgO interface,\cite{Miyakawa2013a} we consider only the linear region of the curve during the fit. The calculated values, given in Table \ref{tbl:summary}, demonstrate that the absence of a Ta insertion leads to the lowest value of $K_i$ ($0.91 \pm 0.09 \mathrm{mJ/m^2}$), explaining why $n$ = 0 samples did not exhibit a perpendicular easy axis. On the other hand, $K_i$ for $n$ = 1 ($2.18 \pm 0.08 \mathrm{mJ/m^2}$) and $n$ = 2 ($2.4 \pm 0.1 \mathrm{mJ/m^2}$) are both larger than the single-MgO series ($1.61 \pm 0.07 \mathrm{mJ/m^2}$), as would be expected from the additional PMA from the second CoFeB/MgO interface. However, the anisotropy \textit{per interface} did not double with the additional CoFeB/MgO, which may be attributed to the different degrees of crystallization for single- and double-MgO samples. An indication of better crystallization into CoFe in the single-MgO series is its higher $M_S$. $K_v$ is negative and does not vary appreciably in samples where Ta is present, in contrast to the positive value found for zero-insertion samples. The role of Ta with regard to $K_v$ is not yet understood, as previously pointed out by Sinha et al. \cite{Sinha2013a}, and a detailed study of the amount, proximity, and profile of Ta would be necessary to clarify these effects.

Turning our attention to $\alpha$, we identify a single-MgO sample ($t_{eff} \approx 0.8$ nm) and a single-insertion sample ($t_{eff} \approx 1.3$ nm) with a comparable $K_{eff}t_{eff} \approx 0.2 \mathrm{mJ/m^2}$. We immediately notice that $\alpha$ for the single-insertion sample is around two times lower than that for the single-MgO sample. 

This dramatic decrease in $\alpha$ may be attributed to the suppression of spin pumping by the MgO layers straddling both sides of the precessing magnet.\cite{Konoto2013,Mosendz2010} Indeed, measurements of zero-insertion samples show no thickness dependence [purple dashed line in Fig.~\ref{fig:Keffalpha}(b)] and a low mean value of $\alpha = 0.0035 \pm 0.0002$ comparable to the bulk damping of $\mathrm{Co_{40}Fe_{40}B_{20}}$\cite{Liu2011b,Oogane2006a}. However, a decrease in $\alpha$ with increasing $t_{eff}$ can still be seen in both the single- and double-insertion series. One possible reason is the alloying of CoFeB and Ta, as Ta is known to readily intermix with CoFeB \cite{Wang2006b}, and higher damping may be expected from CoFeBTa alloys \cite{Rantschler2007}. The relative percentage of CoFeBTa alloy decreases with increasing CoFeB thickness, coinciding with the $\alpha$ decrease. This picture is also consistent with the jump in $\alpha$ from single- to double-insertion samples, i.e., there is more CoFeBTa alloy because there are more Ta insertions. \cite{Tsunegi1882,Konoto2013} 

It may also be possible that spin pumping to the Ta insertion layer occurs, as in the case of the Pd interlayer in CoFe/Pd multilayers \cite{Shaw2012}. The complexity of our system, however, prevents us from using a simple multilayer model. One reason is that the middle CoFeB layer (in the double-insertion case) may have different properties from the CoFeB layers adjacent to MgO, because CoFeB crystallizes from the MgO interface, \cite{Mukherjee2009} with which the middle CoFeB has no contact. The degree of Ta intermixing also depends on the deposition order and will vary across the structure.\cite{Jang2011} At this point, we cannot discriminate the mechanism behind the damping behavior. It may be worthwhile to study the use of CoFeBTa alloys as interlayers to possibly have more control over the amount and distribution of Ta in the stack.\cite{Tsunoda2012}

In conclusion, we have demonstrated PMA and low damping in double-MgO structures. A thin Ta insertion layer was found to significantly increase the PMA - no perpendicular easy axis was realized in our MgO/$\mathrm{Co_{40}Fe_{40}B_{20}}$/MgO films without Ta - and adding more insertions allowed thicker CoFeB layers to remain perpendicular. However, the maximum $K_{eff}t_{eff}$ in double-MgO samples is comparable only with that of the single-MgO sample for this CoFeB composition.\cite{Sato2012} On the other hand, $\alpha$ for double MgO films increases with the number of insertions but is still lower than that of single MgO films for the entire range. Considering both trends with $n$, we find that the optimal stack in the range of samples we studied is a double-MgO, $n$ = 1 sample with $t_{nom} = 1.75$ nm, which exhibits $K_{eff}t_{eff} = 0.27 \mathrm{mJ/m^2}$ at a low damping value of 0.006. 

\section*{Acknowledgement}
We express gratitude for support from the A*STAR Graduate Academy SINGA Program.

\end{document}